\documentclass[twocolumn,prl,showpacs,multicol,amsmath,amssymb]{revtex4}
\usepackage[dvips]{graphicx}
\usepackage{graphics}
\usepackage{color}
\def\be{\begin{equation}}
\def\ee{\end{equation}}

\def\bi{\begin{itemize}}
\def\ei{\end{itemize}}
\def\bn{\begin{enumerate}}
\def\en{\end{enumerate}}
\def\bea{\begin{eqnarray}}
\def\eea{\end{eqnarray}}

\def\ba{\begin{array}}
\def\ea{\end{array}}
\def\bd{\begin{displaymath}}
\def\ed{\end{displaymath}}

\begin{document}
\title{Factorized ground state for a general class of ferrimagnets }

\author{M. Rezai}
\affiliation{Department of Physics, Sharif University of Technology,
Tehran 11155-9161, Iran}
\author{A. Langari}
\affiliation{Department of Physics, Sharif University of Technology,
Tehran 11155-9161, Iran} \email[]{langari@sharif.edu}
\author{J. Abouie}
\affiliation{Department of physics, Shahrood University of Technology, Shahrood 36199-95161,
Iran}
\affiliation{School of physics, Institute for Research in Fundamental Sciences (IPM), 
Tehran 19395-5531, Iran}

\begin{abstract}
We have found the exact (factorized) ground state  of a general class of ferrimagnets
in the presence of a magnetic field which includes the frustrated, anisotropic and long range
interactions for arbitrary dimensional space. In particular cases, our model
represents the bond-alternating, ferromagnet-antiferromagnet and also homogenous
spin $s$ model. 
The factorized ground state is a product of single particle kets
on a bipartite lattice composed of two different spins ($\rho, \sigma$) which
is characterized by two angles, a {\it bi-angle} state.
The spin waves analysis around the exact ground state show two branch of excitations which
is the origin of two dynamics of the model. The signature of these dynamics is addressed as
a peak and a broaden bump in the specific heat.

\end{abstract}
\date{\today}


\maketitle
{\it Introduction.-}
Spin models are the building blocks of the theory of quantum magnetism and
strongly correlated electron systems. In addition,
they have been considered as an effective model to describe the behavior of a system
in several disciplines. Recently, the implementation of quantum notions in
quantum devices has attracted much attentions both in research labs and
demanding applications like nanotechnology, quantum computation\cite{Niel 00} and
particularly optical lattices\cite{Bren 99}.
Quantum spin models are prototype realization of many relevant properties of
quantum implementation in such devices. Therefore, different aspects of
a quantum phase is of utmost importance for scientists and engineers.
Quantum phases are characterized by the ground state (GS) properties
of the corresponding many body
system.\cite{Sach 00}

Except of a few particular cases such as 1D bond alternating
Heisenberg \cite{Suzu 08}, anisotropic Heisenberg  model (XYZ), XXZ in a longitudinal magnetic field and Ising model in transverse field which are exactly solvable\cite{Taka 99},
the GS of a general spin model is not known. However, at some special values of the
model parameters the quantum correlations are vanishing and the GS is a product of
single particle states. The factorized state (FS) manifests zero entanglement which is
necessary to be identified for reliable manipulating of quantum computing. A FS (unentangled) which is
associated with an entanglement phase transition can be also a quantum critical
point in certain condition which is discussed in this article. This information is
also attractive for the study of quantum phase transitions. Moreover, finding an
exact ground state (as a FS) even at particular values of the parameter space
of a many body spin model leads to the identification of that phase in addition to
more knowledge about the properties of the model close to the factorized
point via implementing an approximate method.

In a seminal work, Kurmann, Thomas and M\"uller \cite{Kurm 82} identified the factorized
state of a homogeneous spin-$s$ XYZ chain at a magnetic field of arbitrary direction.
Factorized GS has been also observed in the two dimensional
lattice through Quantum Monte Carlo simulation in terms of entanglement estimators.\cite{Rosc 04} Recently, Giampaolo,
Adesso and Illuminati \cite{Giam 08} have introduced a general
analytic approach to find the factorized ground states in a homogenous
{\it translational invariant} spin-$s$
quantum spin model for arbitrary long range interaction and any dimensional space.
Their study is based on the single-spin unitary operation and the factorized point
is determined at the position where the associated entanglement
excitation energy becomes zero. More elaborate 
explanations came up with some generalizations quite recently.\cite{Giam 09}
The factorized GS of the dimerized XYZ spin chain in a transverse
magnetic field has been investigated and reported that the factorized point
in the parameter space of the Hamiltonian corresponds to an accidental
ground state degeneracy.\cite{Gior 09}
However, in this article we will present (i) the FS of an inhomogeneous
(ferrimagnetic) spin model which is
composed of two spins ($\rho, \sigma$) in the presence of a magnetic field
on a bipartite lattice with arbitrary long
range interaction and dimensional space, (ii) the Hamiltonian is not necessarily
translational invariant and (iii) the exchange couplings can be competing
antiferromagnetic and ferromagnetic arbitrarily between different sublattices to build
many practical models such as frustrated, dimerized and tetramerized materials.
Moreover, our results recover the previous ones for $\sigma=\rho$
and a particular configuration of the couplings. \cite{Kurm 82,Rosc 04,Giam 08,Gior 09}
In addition, we will address on the existence of two
energy scales which lead to a surprising
dynamics of the model close to the factorizing point and its fingerprint as a double peak
in the specific heat versus temperature.
As an enclosure, the results have been applied to the 1D ferrimagnetic XXZ ($\rho$, $\sigma$) spin chain in the presence of a transverse magnetic field which is realized
as a bimetallic substance. \cite{Koni 90} We will also address the cases
where the factorizing field coincides the critical point. 




{\it Factorized state.-} Let us consider a two sites model which is composed of
two spins $\sigma=\frac12$ and $\rho=1$ with the following
Hamiltonian
\be
\label{2h}
H'=J^x\sigma^x\rho^x+J^y\sigma^y\rho^y+J^z\sigma^z\rho^z+ h'(\sigma_z+\rho_z),
\ee
where $J^{\mu}, \mu=x, y, z$ are the exchange couplings in different
directions and $h'$ is proportional to the magnetic field. We are looking for a
factorized state which is satisfied by
 $H'|\sigma\rangle|\rho\rangle=\epsilon|\sigma\rangle|\rho\rangle,$
in which $|\sigma\rangle$ and $|\rho\rangle$ are the single particle states.
It is appropriate to choose $|\sigma\rangle$ and $|\rho\rangle$ to be the
eigenstates of $\vec{\sigma} \cdot \hat{n}'$ and $\vec{\rho} \cdot \hat{n}''$
with eigenvalues $+\frac12$ and $+1$; respectively, where $\hat{n}'(\theta,\varphi)$
and $\hat{n}''(\beta,\alpha)$ are unit vectors in Bloch sphere.
The solution of  $H'|\sigma\rangle|\rho\rangle=\epsilon|\sigma\rangle|\rho\rangle$
gives the factorized state at $h'=h'_f$ and its corresponding
energy ($\epsilon$) [\onlinecite{extension}].
Moreover, we found that the angles $\theta$ and $\beta$ are fixed by the
couplings ($J^{\mu}, h'$; see Eq.(\ref{tbhe})) while $\alpha$ and $\varphi$ are given by one of
these choices (I) $\alpha=0, \varphi=0$, (II)  $\alpha=0, \varphi=\pi$,
(III) $\alpha=\frac{\pi}{2}, \varphi=-\frac{\pi}{2}$,
(IV) $\alpha=\frac{\pi}{2}, \varphi=\frac{\pi}{2}$.
The spins are located in the xz-plane for choices I and II while they have
projections only in the yz-plane for III and IV. Without loss of generality
we can assume the spins are located in the xz-plane. In fact, the spins of yz-plane 
will fall to xz-plane by interchange of $J^x \leftrightarrow J^y$. Moreover, the
coordinates $(\theta,\varphi=0)$ and $(-\theta,\varphi=\pi)$ are representing the same direction, therefore case (I) $\alpha=0, \varphi=0$ is able to describe all possibilities.

The two spin model ($\sigma=\frac12, \rho=1$) is now generalized to
arbitrary ($\sigma, \rho$) spins.\cite{Kurm 82} To find the factorized state of a general
two site ferrimagnet we consider a rotation on $\sigma$ and $\rho$ spins such that
$\overrightarrow{\sigma}$ and $\overrightarrow{\rho}$ point in 
($\theta, \varphi=0$) and ($\beta, \alpha=0$) directions, respectively,
 The rotation
operator is $D= D^{\sigma}(0,\theta,0)D^{\rho}(0,\beta,0)$ where
$D^{\rho}(0,\beta,0)=D(\alpha=0,\beta,\gamma=0)=D_{z}(\alpha)D_{y}(\beta)D_{z}(\gamma)$
is defined in terms of Euler angles and a similar expression is considered for
$D^{\sigma}(0,\theta,0)$.
Then, We impose the condition to have a factorized (fully polarized) eigenstate for this Hamiltonian which
fixes the following relations for the model parameters
\bea
\cos\theta&=&-\frac{h'^2_f J^y+J^x(J^{z^{2}}-J^{y^2})\rho \sigma +h'_fJ^z(J^y \rho+J^x \sigma)}{h_f^{'2} J^x+J^y(J^{z^{2}}-J^{x^2})\rho \sigma +h'_f J^z(J^x \rho+J^y \sigma)}, \nonumber\\
\cos\beta&=&-\frac{h'^2_f J^y+J^x(J^{z^{2}}-J^{y^2})\rho \sigma +h'_fJ^z(J^y \sigma+J^x \rho)}{h'^2_f J^x+J^y(J^{z^{2}}-J^{x^2})\rho \sigma +h'_f J^z(J^x \sigma+J^y \rho)}, \nonumber \\
h'_f&=& \sqrt{\frac12\big(2J^x J^y \rho \sigma+(\rho^2+\sigma^2)J^{z^2}+C J^z\big)}, \nonumber\\
C&\equiv&\sqrt{4\rho\sigma(\rho J^x+\sigma J^y)(\sigma J^x+\rho J^y)+(\rho^2-\sigma^2)^2 J^{z^2}}, \nonumber\\
\epsilon&=&\frac{J^x J^y}{J^z}\sigma\rho-\frac{h'^2_f}{J^z}.
\label{tbhe}
\eea
Therefore, for arbitrary ($\sigma, \rho$) and at the above value for $h'=h'_f$ we have a
fully polarized eigenstate which is a factorized state. The ordering of this state is defined by
two angels ($\theta, \beta$) which show the orientations of
($\overrightarrow{\sigma}, \overrightarrow{\rho}$), respectively.

Now, we intend to find the condition for having a factorized state for a ferrimagnetic
lattice in a magnetic field. We will then show that the factorized state is the ground state
of lattice implementing a spin wave theory to study the quantum fluctuations. We consider a general Hamiltonian of ferrimagnets
on a bipartite lattice where sublattice ($A_{\sigma}$) contains $\sigma$ spins
and the other sublattice ($B_{\rho}$) includes
$\rho$ spins.
The interaction can be long ranged between different sublattices but no interaction in the same sublattice.
The ferrimagnetic Hamiltonian for such case can be written as
\begin{eqnarray}
\label{m.b.h}
H&=&\sum_{i,r}\left[ \zeta_i \hat{\zeta}_{i+r} 
(J^x_r\sigma^x_{i}\rho^x_{i+r}+J^y_r\sigma^y_i\rho^y_{i+r})+
J^z_r\sigma^z_i\rho^z_{i+r}
\right] \nonumber \\
&+&h \sum_{i}(\sigma^z_i+\rho^z_{i}),
\end{eqnarray}
where $i=(i_1, i_2, i_3)$ and $r=(r_1, r_2, r_3)$ are representing
the three dimensional
index on the lattice and $\zeta_i, \hat{\zeta}_{i+r}=\pm1$ which realize both ferromagnetic (F)
and antiferromagnetic (AF) exchange interactions. 
A remark is in order here, the Hamiltonian in Eq.(\ref{m.b.h})
is a sum of two sites Hamiltonian defined in Eq.(\ref{2h}) where the two spins
can be far from each other. However, the interaction between each couple of
($\sigma_i, \rho_{i+r}$) can depend on distance ($r$) with different strength and
also be F or AF arbitrarily defined by $\zeta_i, \hat{\zeta}_{i+r}$.
A factorized eigenstate for the Hamiltonian of Eq.(\ref{m.b.h})
can be written as
\be
\label{fs}
|FS\rangle=\bigotimes_{i\in A_{\sigma}, j\in B_{\rho}}|\sigma'_{i}\rangle
 |\rho''_{j}\rangle
\ee
where $|\sigma'_{i}\rangle$ and $|\rho''_{j}\rangle$ are the eigenstates of
$\vec{\sigma}_i\cdot\hat{n}_i'$ and $\vec{\rho}_j \cdot\hat{n}_j''$
with largest eigenvalue where $\hat{n}_i'$ and $\hat{n}_j''$ are
unit vectors pointing in ($\zeta_i \theta, \varphi=0$) and
($\hat{\zeta}_j \beta, \alpha=0$), respectively.
However, the factorized state ($|FS\rangle$)
is an eigenstate of the Hamiltonian
if the angle $\zeta_i \theta$ ($\hat{\zeta}_j \beta$) be consistent
with all pair of interactions originating from $\sigma_i$ ($\rho_i$) on
sublattices $A_{\sigma}$ ($B_{\rho}$). According to Eq.(\ref{tbhe}) the former condition is
satisfied if the interaction between each pair ($\sigma_i, \rho_{i+r}$) is
the same for all directions while depending on distance ($r$), i.e
$J^{\mu}_r=\lambda(r)J^{\mu}, \mu=x,y,z, \lambda(r)>0$. Under these constraints the
factorized state (Eq.(\ref{fs})) is an eigenstate of $H$ with the characteristic
angles ($\theta, \beta$) defined in Eq.(\ref{tbhe}) and the factorizing field is 
\be
\label{hf}
h_f= h'_f \sum_{r=0}^{N_r} \lambda(r),
\ee 
where $N_r$ is the number of spins on each
sublattice. 
To show that $|FS\rangle$  is the ground state of $H$ at $h_f$ we
first implement a rotation on the Hamiltonian.
The rotated Hamiltonian ($\tilde{H}$) is the result of
rotations on all lattice points of $H$,
\be
\tilde{H}= \tilde{D}^{\dag}H\tilde{D}\;,\;
\tilde{D}=\bigotimes_{i\in A_{\sigma}, j\in B_{\rho}}
D_i^{\sigma}(0,\zeta_i \theta,0)D_j^{\rho}(0,\hat{\zeta}_j \beta,0).
\ee
In the next step the rotated Hamiltonian is
bosonized using the Holstein-Primakoff (HP) transformation, 
$\sigma^{+}_i=\sqrt{2\sigma-a^{\dag}_ia_i} \ a_i,
\sigma^{z'}_i=\sigma-a_i^{\dag}a_i,
\rho^{+}_j=\sqrt{2\rho-b^{\dag}_jb_j} \ b_j,
\rho^{z''}_j=\rho-b^{\dag}_jb_j,$
where $a_i (a^{\dagger}_i)$ and $b_j (b^{\dagger}_j)$ are two types annihilation (creation)
boson operators.
Using HP transformations the anisotropic ferrimagnetic spin model is mapped to an interacting system of bosons.
The Hamiltonian in the momentum ($k$) space and in the linear spin wave
theory is diagonalized via the rotation, $\chi_{k}=a_{k}\cos\eta_{k}-e^{i \delta} b_{k}\sin\eta _{k};
\psi_{k}=e^{i \delta} b_{k}\cos\eta_{k}+a_{k}\sin\eta_{k}$, and a shift at $k=0$ where
$\delta$ is defined by $\sum_{r} e^{-i k \cdot r} J_r^y=|\sum_{r} e^{-i k \cdot r} J_r^y| e^{i \delta}$. The diagonalized
Hamiltonian is
\be
\tilde{H}=E_{gs}+\sum_{k}\bigg(\omega^{-}(k)\chi_{k}^{\dag}\chi_{k}+
\omega^{+}(k)\psi_{k}^{\dag}\psi_{k}\bigg),
\ee
where $E_{gs}$ is the ground state energy \cite{extension} and
$\omega^{\pm}(k)\geq0$ are normal modes parallel and perpendicular to the field direction.
\bea
\omega^{\pm}(k)&=&D^+\pm\frac{D^{-}+\sqrt{\sigma \rho}\tan(2\eta_k)
|\sum_{r} e^{-i k \cdot r} J_r^y|}{\sqrt{1+\tan^2(2\eta_k)}}, \nonumber \\
&&\tan(2\eta_{k})=\frac{\sqrt{\rho\sigma} |\sum_{r} e^{-i k \cdot r} J_r^y|}{D^-}, 
\eea
in which
\bea
D^{\pm}&\equiv&\frac{h^2_{f}}{2\tau^z}(\frac{1}{\rho}\pm\frac{1}{\sigma})+h_f(\frac{\sigma}{2\rho}\cos\theta\pm\frac{\rho}{2\sigma} \cos\beta )\nonumber \\
&-&\frac{\tau^x \tau^y}{2\tau^z}(\sigma\pm\rho)+\frac{h_f-h}{2}(\cos\beta\pm\cos\theta)
, \nonumber \\
&&\tau^{\mu}=J^{\mu} \sum_r \lambda(r).
\eea
 The bosons number $\langle a^{\dag}a\rangle$ is 
proportional to $(h_f-h)$
which states that at $h=h_f$ the bosons number in the
ground state is exactly zero \cite{extension}, i.e 
$\langle a^{\dag}_i a_i\rangle|_{h=h_f}=0=\langle b^{\dag}_i b_i\rangle|_{h=h_f}$.
It is also true for 
$\langle a^{\dag}_i b_i\rangle|_{h=h_f}=0$.
Thus, the spin wave theory is exact in all orders
at the factorizing field ($h=h_f$) and $|FS\rangle$ is the corresponding ground state
of $H$.

\begin{figure}
\begin{center}
\includegraphics[width=8cm]{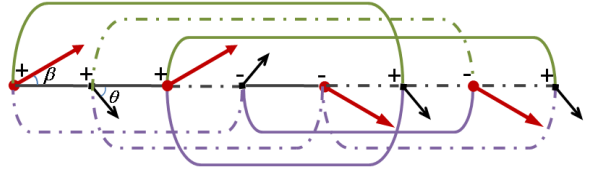}
\caption{The configuration of a factorized state on a one-dimensional lattice
for arbitrary frustrated couplings. Solid lines (dash-dotted)
represent antiferromagnetic (ferromagnetic) couplings which are defined
by $\zeta_i \hat{\zeta}_{i+r}$ as depicted by $\pm$ on each site.
Each color belongs to equal distance interaction (same $r$).
} \label{chain}
\end{center}
\end{figure}

To visualize the configuration of a factorized ground state of a general
frustrated model we have plotted an example in Fig.(\ref{chain}) with the assumption
$J^{x}_r, J^{y}_r>0$ and $J^{z}_r<0$ where $\zeta_i$ and $\hat{\zeta}_{i+r}$ define
the sign of interactions. The solid-lines
represent antiferromagnetic interaction and the dash-dotted ones are the
ferromagnetic counterparts. As shown in Fig.(\ref{chain}) the interactions can
be frustrated and long ranged without a translational invariance.
However, the factorized state is defined by two angles ($\theta, \beta$) while
each $\sigma$ ($\rho$) spin is directed in $\theta$ ($\beta$) or
$-\theta$ ($-\beta$) directions. 
We call this a {\it bi-angle} ordering. In a special case
the bi-angle ordered state can configure a ferromagnet ($\theta=\beta$) or 
antiferromagnet ($\theta=\pi-\beta$) factorized state.


{\it Discussions.-}
In a spin model when the magnetic field is strong enough all spins will align in the
direction of the magnetic field which characterizes the saturated phase as
far as $h\geq h_s$. In our notation, the saturated phase appears when 
all $\sigma$ ($\rho$) spins
get $\theta=\pi$ ($\beta=\pi$). 
Thus, the saturating field ($h_s$) is a factorizing one when $J^x=J^y$.
In case of $J^x \neq J^y$ the saturation can only appear at infinite value of the
magnetic field while a finite factorizing point ($h_f$) still exists.
For $J^x=J^y$, the lower
excitation band  becomes gapless ($\omega^-(k=0)=0$) which confirms that
the factorizing point ($h_f=h_s$) is the critical point which separates the non-saturated
phase ($h<h_s$) from the saturated one ($h>h_s$).
It is worth to mention that at $J^x=J^y$ the rotational symmetry around the magnetic
field is restored where the quantum fluctuations around the field axis are suppressed.
In the absence of rotational symmetry the saturation in the field direction does not
happen at a finite value of magnetic field while the model approaches saturation 
asymptotically at infinity.

A general feature of our result is that it can simply recover the previous study
of homogenous systems by replacing $\sigma=\rho=s$. In that case the restriction of
bipartite lattice is promoted to arbitrary lattice and the interaction between any pair
of spins can exist. 
However, our Hamiltonian is not restricted to the
translational invariant symmetry or bond-alternating ones which is witnessed by the
example given in Fig.(\ref{chain}). This can also be generalized to any dimension.
We claim that the general Hamiltonian which can possess a nontrivial 
factorized ground state should
be of the form Eq.(\ref{m.b.h}) with the restriction $J^{\mu}_r=\lambda(r)J^{\mu}, \mu=x,y,z$.

In the absence of magnetic field, the spin-($\rho, \sigma$) ferrimagnet turns into 
the spin-$\sigma$ antiferromagnet in the limit $\rho \rightarrow \sigma$, whereas it 
looks like the spin-$\rho$ ferromagnet in the other extreme limit 
$\frac{\rho}{\sigma} \rightarrow \infty$. In this sense, the difference $\rho-\sigma$
can be regarded as the ferromagnetic contribution. The analysis of our result shows
that the value of $h_f$ is a decreasing function of $\frac{\rho}{\sigma}$ converging
to the ferromagnet feature as this ratio becomes large. More investigations on different 
applications of our approach in several models are in progress.\cite{extension}

Let us now be more concrete by concentrating on the one-dimensional nearest neighbor
XXZ ferrimagnet in the presence of transverse magnetic field.
Suppose that $J^x=J^z=J$ and $J^y=J \Delta$ where $\Delta$ represents the easy axis
anisotropy.
At zero magnetic field the quantum fluctuations
are large and the ground state of the model is strongly entangled. Upon adding the transverse magnetic field the U(1) symmetry of the XXZ model is lost and the entanglement of the GS is decreased. In the mapped bosonic system
the magnetic field is served as a chemical potential, thus the number of bosons ($\langle a^{\dagger}a\rangle$ and $\langle b^{\dagger}b\rangle$)
is dependent on the magnetic field. An enchantment of the magnetic field causes deducing of the bosons' number and the quantum correlations decrease.
At factorizing field $h=h_f$, the number
of bosons is zero and the quantum fluctuations become completely uncorrelated.
Moreover, our calculations show that the factorizing
field in ferrimagnetic model depends on the anisotropy parameter ($\Delta$) similar 
to a homogeneous
antiferromagnetic Heisenberg model.
For $\Delta=0$ (ferrimagnetic XY model) the in-plane magnetic field completely breaks the rotational symmetry. Increasing $\Delta$ suppresses the effect of the field and try to evoke the rotational symmetry to the system. Thus, by increasing $\Delta$ the factorizing Ne\'{e}l field approaches to the saturation field.
At $\Delta=1$, the rotational symmetry is completely repayed and $h_f$ is exactly lied on $h_{s}$.

\begin{figure}
\centerline{\includegraphics[width=7cm,angle=0]{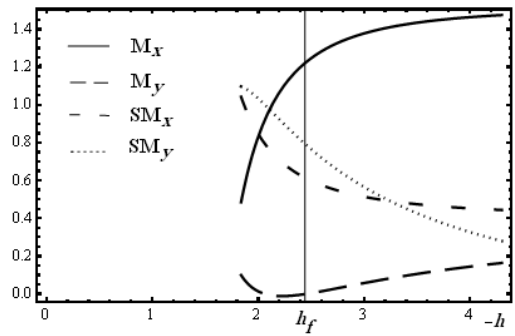}}
\caption{The magnetization and staggered magnetization of an anisotropic ferrimagnetic
($\sigma=1/2, \rho=1$) spin chain versus transverse field and for $\Delta=0.25$.}
\label{magnetization}
\end{figure}

A benefit of identification a factorized state is that we can work out an approximate method
around the factorized point to get some information on the properties of that phase.
This help us to calculate the magnetic properties of the ferrimagnetic XXZ model
in the presence of a transverse magnetic field. We have implemented the linear spin wave
theory around $h=h_f$ for $\sigma=\frac{1}{2}$  and $\rho=1$. Our results for magnetization
($M_x, M_y$) and staggered magnetization ($SM_x, SM_y$) in both $x$ and $y$ directions
are plotted in Fig.(\ref{magnetization}) where the magnetic field is in $x$ direction.
The approach can be extended to derive the thermodynamic of this model which will be
done in near future.

It is also worth to mention that our results are applicable to the homogenous XXZ Heisenberg spin-1/2 chains in the presence of a transverse magnetic field ($h^x$). This model has been studied intensively in the literature.\cite{Kenz 02, langari 04, Siah 08, Abou 08}
On the onset of transverse magnetic field a perpendicular anti-ferromagnetic 
order is stabilized by promoting a
spin-flop phase. At the classical (factorizing) field ($h_f<h_c$) in the spin-flop phase the ground
state is factorized to single spin states and the
staggered magnetization along the $y$ direction has a large value. Very close to the critical field ($h_c$) the anti-ferromagnetic order becomes
unstable and the staggered
magnetization falls sharply to vanish at the critical point.
For $h>h_c$ the spins are almost aligned in field ($x$) direction and 
the factorized state as a fully polarized phase will be appeared for $h\rightarrow \infty$.
The excitation energies around the factorizing point have the following form:
\begin{eqnarray}
\omega^{\pm}(k)=(1+\Delta)\frac{h}{h_f}+\Delta(\pm \cos(\frac{k}{2})-1).
\end{eqnarray} 
Thus, we have two branches of magnon energies as two scales which impose two dynamics in the system.
The most interesting feature is that around the factorizing field both scales show up.
 These dynamics correspond to the coexistence of
two different features of the model. In other words, by increasing the
transverse magnetic field, spins try to align in the $x$ direction which is the
ferromagnetic feature of the system. In the intermediate values of
$h$, we have both anti-ferromagnetic and ferromagnetic behaviors since the
anti-ferromagnetic order in $y$ direction has already been stabilized.
The finger print of these features appear in the thermodynamic functions such as specific heat and internal energy. As it is seen from Ref.[\onlinecite{Abou 08}] the second feature can be seen as a shoulder at the
right side of specific heat curve. By further increasing of $h$,
the ferromagnetic behavior is seen as a broaden peak in the curve.
This point is almost near the classical field where the ground state
of the system has been factorized.

This work was supported in part by the Center of Excellence in
Complex Systems and Condensed Matter (www.cscm.ir).




\begin{thebibliography}{99}
\bibitem{extension}
More details of the calculations will appear elsewhere.
\end{thebibliography}
\end{document}